\documentclass[12pt, draftclsnofoot, onecolumn]{IEEEtran}
\usepackage{cite,graphicx,amsmath,amssymb}
\usepackage{subfigure}
\usepackage{citesort}
\usepackage{fancyhdr}
\usepackage{mdwmath}
\usepackage{mdwtab}
\usepackage{balance}
\usepackage{xcolor}
\usepackage{bm}

\usepackage{algorithm}
\usepackage{algorithmic}
\usepackage{multirow}
\usepackage{flafter}

\usepackage{multicol}
\usepackage{amsthm}
\usepackage{graphicx}
\usepackage{amsmath}
\usepackage{flafter}

\newtheorem{theorem}{Theorem}

\begin{document}


\title{Deep Learning for Latent Events Forecasting in Twitter Aided Caching Networks}
\author{Zhong\ Yang,~\IEEEmembership{Student Member,~IEEE,}
Yuanwei\ Liu,~\IEEEmembership{Senior Member,~IEEE,}\\
Yue\ Chen,~\IEEEmembership{Senior Member,~IEEE}, and
Joey\ Tianyi\ Zhou,~\IEEEmembership{Senior Member,~IEEE}

\thanks{ Part of this paper has been presented in IEEE Global Communication Conference (GLOBECOM) 2019~\cite{Qi2019globecom}.}
\thanks{ Z. Yang, Y. Liu and Y. Chen are with the School of Electronic Engineering and Computer Science, Queen Mary University of London, London E1 4NS, UK. (email:\{zhong.yang, yuanwei.liu, yue.chen\}@qmul.ac.uk)}
\thanks{ Joey\ Tianyi Zhou is with Agency for Science Technology and Research (A*STAR), Singapore. (email:~joey.tianyi.zhou@gmail.com)}
}
\maketitle

\begin{abstract}
A novel Twitter context aided content caching (TAC) framework is proposed for enhancing the caching efficiency by taking advantage of the legibility and massive volume of Twitter data. For the purpose of promoting the caching efficiency, three machine learning models are proposed to predict latent events and events popularity, utilizing collect Twitter data with geo-tags and geographic information of the adjacent base stations (BSs). Firstly, we propose a latent Dirichlet allocation (LDA) model for latent events forecasting taking advantage of the superiority of LDA model in natural language processing (NLP). Then, we conceive long short-term memory (LSTM) with skip-gram embedding approach and LSTM with continuous skip-gram-Geo-aware embedding approach for the events popularity forecasting. Lastly, we associate the predict latent events and the popularity of the events with the caching strategy. Extensive practical experiments demonstrate that: (1) The proposed TAC framework outperforms conventional caching framework and is capable of being employed in practical applications thanks to the associating ability with public interests. (2) The proposed LDA approach conserves superiority for natural language processing (NLP) in Twitter data. (3) The perplexity of the proposed skip-gram based LSTM is lower compared with conventional LDA approach. (4) Evaluation of the model demonstrates that the hit rates of tweets of the model vary from 50\% to 65\% and the hit rate of the caching contents is up to approximately 75\% with smaller caching space compared to conventional algorithms.
\end{abstract}

\section{Introduction}\label{section:introduction}

Recent advances in mobile smart devices, ubiquitous social media and application brings tremendous expansion of mobile data traffic. The visual networking index (VNI) report from Cisco~\cite{Cisco2019} reveals that global mobile data traffic (GMDT) is expected to grow to 77 exabytes per month by 2022, a seven-fold increase over 2017. Specifically, 5G will be 3.4 percent of connections but 11.8 percent of total traffic by 2022. The extensive GMDT growth and innovative network technology compel network providers to investigate novel techniques for sufficing the network services and easing up backhaul transmission.

Content caching at networks edges is a prospective approach for reducing the network backhaul transmission and bringing down the network delay~\cite{Ioannou2016ICST,Tang2020maga}. Nevertheless, caching superiority is related closely with the popularity of the contents in the network. In conventional caching frameworks, the user preference is assumed following a generalized Zipf law~\cite{Wang2017twc}: the content request rate $r\left( i \right)$ for the $i$-th content in the network is proportional to ${1 \mathord{\left/
{\vphantom {1 {{i^\alpha }}}} \right.
\kern-\nulldelimiterspace} {{i^\alpha }}}$ where $\alpha$ is the temperature of the network and typically less than Unit. However, the Zipf law is a experience distribution and lacks theoretical guarantee. Specifically, the counting methods based on Zipf distribution properties only demonstrate the frequency of the words rather than the concurrency between the words. Therefore, the extracted topics are incomplete, which impairs the performance of text-related content prediction in wireless caching. Moreover, employing the properties of Zipf law of text-related content to determine what to cache regardless of the locations of base stations (BS) is pervasive yet prodigal~\cite{twenty_749260}, especially for the legibility and massive volume of social media data. According to~\cite{Statista2019}, social media usage is one of the most popular online activities and the number of people using social media worldwide increases to almost 3.43 billion, 3.5 times that of 2010.

Social media motivated caching strategy has attracted attentions from both academia and industry~\cite{Tsai2020tnse,Xiao2020Inet,Zhu2017INET,Zeydan2016ICOMNET}. The authors in~\cite{Tsai2020tnse} analyse the Twitter data of 2016 U.S. presidential election utilizing LSTM networks to reduce the service latency. A preference-aware optimization~\cite{Xiao2020Inet} considers user side adaptive streaming, coordinated bandwidth allocation, and network side caching content selection. In~\cite{Zhu2017INET}, caching cost of the base stations (BS) and social factors among mobile users (MU) are considered in ultra-dense small cell networks (UDN) to obtain effective caching incentives and the optimal social group utility. Zeydan, etal,~\cite{Zeydan2016ICOMNET} proposed a big data enabled caching architecture, in which a vast amount of data is harnessed for content popularity estimation and content caching. Twitter is one of the most popular social media platforms in the world, that contains countless open accessed tweets (messages) published by the users from different regions. With billions of new tweets being posted every day, the freshness and chronological variation of the text contents in tweets are attracting more and more researches to exploit tweets to gather large amounts of public data for big data related researches.

As the public tends to post their preferences and interests on social media platforms, it brings us an opportunity to cache text-related content more accurately in the BS through topics/events prediction. To efficiently predict caching text-related contents among different BS, it is plausible to associate the public preference with the Twitter topical issues. Top words in Twitter topics have been proved to vary according to different locations in London \cite{three_lansley2016geography}. Twitter data indicate the preference of the public towards specific topics \cite{two_o2010tweets}. After extracting latent topics from tweets, the text-related contents caching in the BS can be determined. However, the Zipf-distribution is not sufficient accurate since the structure of the natural language shows a statistic structure beyond Zipf distribution properties\cite{twenty-one_piantadosi2014zipf}, which leads to incapability of Zipf-based algorithms in extracting latent topics/events from tweets..


Big data techniques have enabled the industry to deal with large amounts of data in high efficiency.  Wireless content caching system is the crucial key to improve the efficiency of the wireless caching at the edge of the wireless network. However, it is always nontrivial to determine what to cache at the wireless devices\cite{twenty-two_6195469}. With the aid of friendly application programming interface (API) offered by the social media platforms, we are able to easily  filter the social media contents of large amounts within the constrained region. Since the APIs also enable users to filter the regions of tweets based on their location tags (geo-tags), relating regional preference to the tweets in that region seems to be a feasible way to determine what to cache based on the public preferences. Combining the APIs with big data techniques, individual BS is therefore capable of automatically allocate the local public-posted information. As aforementioned, with the aid of the machine learning approaches, the preference is predicted and the text-related caching contents are associated with the preference of the regional public.

\subsection{Prior Work}
Caching at the edge of the wireless networks is considered one of the most important direction due to its great potential of forwarding desired contents during the rush hours, which alleviates the network burden~\cite{twenty-four_8403956}. With the increased number of mobile users in wireless networks, pushing frequently requested contents close to them has been deemed as a efficient way of overcoming the bottlenecks of the wireless communication systems. Furthermore, caching contents according to the preference of the MUs can improve the efficiency and prevent congestions to some extent~\cite{twenty-two_6195469}. According to~\cite{twenty-five_6495773}, video content caching approaches have been proved to be efficient in improving video throughput.

Machine learning approaches and social media platforms have attracted more attention in the field of wireless caching recently. With the aid of machine learning approaches, proactive allocating systems are able to better predict the wireless traffic patterns\cite{twenty-three_6871674}. By exploiting topical issues in the regions and the interactions among the public, the networks are capable of better predicting the text-related contents at the edge of the networks. To reinforce the quality of the wireless network service, caching has been proved to be effective in caching radio contents\cite{twenty-six_pangtoward} and the hit rate has been improved.

Besides, with the aim to establish a proactive device-to-device (D2D) caching network, authors of a paper linked the users together to share caching contents\cite{twenty-three_6871674}. Aiming to extract contextual contents from users' interactions, there is also a paper proposed a framework of wireless caching\cite{twenty-eight_bacstuug2015transfer}.
According to the prior works, the application of machine learning approach has been applied on the video caching field to reinforce the wireless caching network \cite{twenty-six_pangtoward}.
There is also a paper which employs several parameters to evaluate the different candidate contents through making evaluations on their popularity profile\cite{twenty-seven_6883600}.

Related works include the effort to associate different users of the social networks with the aim to establish a proactive D2D caching network\cite{twenty-three_6871674}. The thought is to retrieve the cached contents from other users to satisfy the requests from others users for the same contents. Similar to this work, there is also a paper proposed a framework of wireless caching\cite{twenty-eight_bacstuug2015transfer}, which extracting the contextual information from the users' interactions. The results mentioned in the above two paper demonstrate that the contextual extraction wireless caching models are capable of reinforcing the accuracy of wireless caching and reducing the redundancy of the BS caching contents.
Acknowledged from the prior works, the application of machine learning approach has been applied on the video caching field \cite{twenty-six_pangtoward}. In this paper, the authors demonstrated several approaches towards determining the video contents to be cached at the BS, including Least Recently Used (LRU), Least Frequently Used (LFU) and their proposed machine learning approach. The hit rates of their models vary from 80\% to 90\% as the caching size varies from 10 GB to 100 GB. However, in their model, the monotonic video caching dataset is exploited, which related poorly to the text and other aspects of public preference.
There is a paper exploits several parameters to determine the popularity of the candidate contents to be cached\cite{twenty-seven_6883600}. In this paper, authors classify the contents based on their "popularity profile", which is determined by algorithms. After the evaluation, the data with high popularity profile are cached at the BS while the other are downloaded from the network. Final result demonstrated that the fetch rate increases along with the popularity profile while the fetch rates vary from 15\% to 25\% based on their proposed algorithm.

Therefore, based on the two prior papers mentioned above, we are capable of concluding that the prediction and determination of complex, comprehensive wireless caching contents, which including different types of data (video, images, text, etc.) are increasingly more nontrivial compared to the wireless caching model towards one monotonic caching content. Further, deep learning approaches have been proved to be effective and efficient while determining wireless caching contents.


\subsection{Motivation and Contribution}
Our motivation is to combine different types of machine learning models to propose a wireless caching framework, which exploits social media data as reference to determine the text-related contents at BS. Since there are different data types in the twitters including images, videos, music, etc., the proposed model is general and feasible of multiple caching context. As mentioned above, social media platforms have been regarded as a major network traffic consumer due to their popularity\cite{fourteen_yang2016estimating}. After determining what heavy-weighted caching contents (images, videos) to cache, the caching-enabled wireless BS is capable of reducing the burden of the network and reduce backhaul capacity.
Since allocating different topics of text-related caching content requires distinct parameters of Zipf-based models in different areas to obtain satisfying caching predictions, the efficiency of setting up such networks is limited and the interactions among the networks are limited.
Therefore, an autonomous and reliable wireless caching framework is desired. Furthermore, when indigenous BSs are capable of forming a autonomous region to reinforce multi-BS caching, which leads to less computing costs and wastes.

One core problem is that tweets are considered to be challenging for topics allocating due to their colloquialism, short length (less than 280 components) as well as the informal usage of language \cite{five_ramage2010characterizing}. The conventional counting approaches, like latent semantic analysis (LSA)~\cite{Wang2016sple}, are insufficient to find the topics. To enhance the accuracy, we try to apply the following two machine learning models, namely latent Dirichlet allocation (LDA), and long short-term memory (LSTM). The Beyes topic modelling approach, latent Dirichlet allocation (LDA), which employ multinomial probability over terms for topic allocation \cite{six_steyvers2007probabilistic}. LDA is able to obtain precise outcome from data of social media like Twitter \cite{eight_fang2016topics}. As the preference of the public in the same region tends to vary chronologically, we employ long short-term memory (LSTM) \cite{nine_hochreiter1997long} to model long-term contextual information. LSTM is an approach of recurrent neural network (RNN), which is feasible of allocating topics based on context\cite{fifteen_sutskever2014sequence}.


The other core problem is how to associate the tweet text, which represents the public preference, with the solid caching contents. In an effort to associate Twitter topics with corresponding BS, we propose the approach to arrange the tweets to their pertinent BS in London. Previous studies on Twitter demonstrate that Twitter topics with geographic information represent the preference of the public. After obtaining the topics from the tweets, we associate the topics with actual tweet text to determine what media files to cache in order to achieve better performance. Considering the difference between that the caching background of social media platforms and that of traditional content streams (i.e. the video sites), we propose criteria to demonstrate the coherency between the topics and future media contents. Our contributions are summarized as follows:
\begin{itemize}
  \item We propose a novel Twitter context aided content caching (TAC) framework in which Twitter topics with BS geography information are associated with social media platform APIs. In this practical framework deep learning approaches are conceived in each BS to reinforcing the autonomous determination of wireless text-related contents caching at the edge of wireless networks.

  \item We conceive a versatile LDA model for latent events forecasting utilizing collected practical data from Twitter. The proposed LDA model is capable of conserving superiority for natural language processing (NLP) in Twitter data.

  \item We adopt a LSTM with skip-gram embedding for content popularity forecasting. In an effort to predict latent events as well as their locations, we propose a new model that exploits LSTM and continuous skip-gram-Geo-aware embedding approach to forecast not only the terms of the prospective Twitter events but also the location where the tweet are posted.

  \item Extensive practical experiments verified the performance of the proposed framework and algorithms. The proposed TAC framework is capable of generating satisfactory forecast results among different regions. Judging the caching system with the hit rates upon datasets of different sizes, the hit rates of tweets vary from 50\% - 65\% while the hit rate of the caching contents reaches up to approximately 75\%. The framework also saves caching spaces compared to the conventional algorithms.

\end{itemize}

\subsection{Organization}
The paper is arranged into following sections. In Section \uppercase\expandafter{\romannumeral2}, related works, which employ machine learning approaches to solve wireless caching problems, as well as their performances are listed. In Section \uppercase\expandafter{\romannumeral3}, the structure of the framework and the preparing procedure of the datasets are demonstrated. In Section \uppercase\expandafter{\romannumeral4}, the problems of topics extraction and prediction along with the solutions are demonstrated. In Section \uppercase\expandafter{\romannumeral5}, experiments are proposed to evaluate the model. Numerical and literal results are listed to evaluate the topic extraction procedures. In Section \uppercase\expandafter{\romannumeral6}, the topics obtained from Section \uppercase\expandafter{\romannumeral5} are applied to determine the wireless caching contents. Certain numerical results are demonstrated to present the accuracy and efficiency of our approaches.

\section{TAC Framework and Dataset Preparation}
In this section, the proposed TAC framework and process of Twitter dataset preparation are demonstrated.

\subsection{TAC Framework}

\begin{figure}[htbp]
   \begin{center}
        \includegraphics[width=3.4in]{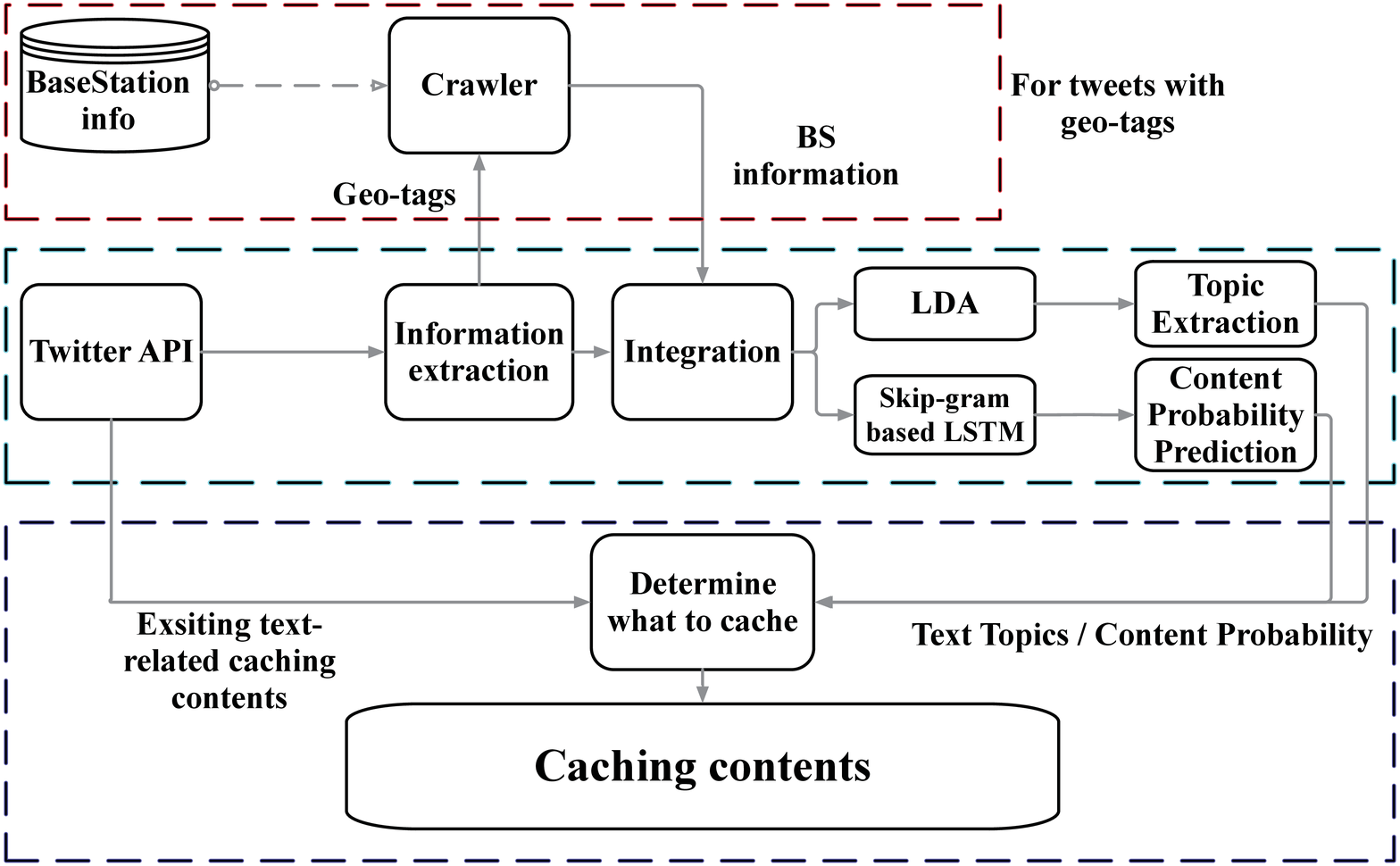}
        \caption{The proposed Twitter aided content caching framework.}
        \label{network}
    \end{center}
\end{figure}

The TAC framework is illustrated in Fig.\ref{network}. The tweets are collected through the Twitter API with their geo-tags. After extracting the geo-tags from the tweets, the information of the nearest BS is obtained through a python Crawler. After offering the crawler with the geo-tags from tweets, the crawler returns the locations of BS\footnote{BS data retrieved from https://mastdata.com}. Afterwards, the dataset is arranged into different regions as an integration of the text with the geographic information. Thereinafter, three machine learning models are invoked to generate predictions according to the collected dataset from Twitter. Finally, since the Twitter API is offering URLs directing to the media files (images, videos, etc.) if the files are contained in a tweet, the framework determines which of the files to cache based on the descriptions in tweets associated with the topics prediction. The dataset throughout the paper was collected through the Twitter API with filter, which gathers the tweets with geo-tags in London. The latitude and the longitude restrictions are given as follows:
\begin{itemize}
  \item Latitude restriction from 51.7136401 to 51.3679144
  \item Longitude restriction from 0.285472 to -0.4488468
\end{itemize}

\subsection{Dataset Preparation}
The Twitter dataset is collected between 27th January 2018 and 27th February 2018 (31 days in total) which composes approximately 70000 tweets with geo-tags in total. Compared with the parallel collecting experiment, which exploits the keyword ``UK" to filter tweets and obtained 2 million tweets in total, the portion of tweets with geo-tags is approximately 2.89\% of the tweets with keyword ``UK". The coordinates of the tweets are restricted by the parameters as demonstrated above. The time range within a day of the collection is constrained from 7:00 AM to 15:30 PM (GMT).

\subsubsection{Distribution of the data}
Due to the different distribution of population and BS in London, the magnitude of the tweets within different districts varies during the same time period as illustrated in the density map. We divide the entire London area equally into nine smaller areas by latitude and longitude of BS locations.


%
\subsubsection{Data cleaning}
To satisfy the prerequisites of training machine learning models with the dataset, some characters are deleted based on the standards below:
\begin{itemize}
  \item Characters that do not belong to English.
  \item Punctuation and Stopwords (in the nltk.corpus package of Python).
  \item Numbers.
\end{itemize}

Since URLs in the tweets are largely points to specific objects, such as a web page or a piece of video, they are valuable in selecting caching contents. The URLs directing to the media files in tweets are stored separately with the tweet text. Therefore, it offers us an opportunity to associate the caching media files with their tweet text.

Particularly, the twitter tags, such as ``\#London", are not removed as they are able to be deemed as constrains of latent topics. The reason that we employ tweet text and the tags rather than merely the tags is because 1) Not all tweets contains tags, some users are not accustomed to use tags. 2) The description of the files is not sufficient with tags---some tags are too vague to describe the media files. For instance, for a particular episode of a TV series, the tag is the whole TV series.

\subsubsection{Caching-contents datasets}
Regarding the fact that when a tweet contains certain text-related media files---images, videos, the URLs directing to the files are given when retrieving the tweets from Twitter API. Since we need to determine the exact caching contents based on tweet text, it is necessary to obtain the caching contents for the text datasets mentioned above.
Among the approximately 70000 tweets with geo-tags in total, 7.69\% of them contain media files(images and videos). As the tweets with videos are approximately 10 times more compared to the ones with videos, images take up a larger proportion in tweets' media files. Furthermore, since the videos in tweets are mainly short videos, the total occupied caching space of images is twice as large as that of videos. According to the reasons above, in this paper, we focus on caching both the images and videos from the tweets rather than merely the videos.

\section{LDA Learning for Latent Events Forecasting}
After demonstrating our thought of the structure of the proposed TAC framework, the main concept of this subsection is to determine text-related caching contents based on twitter events. Regarding our proposed TAC framework, obtaining Twitter topics is necessary for determining the caching contents. Therefore, the methods in this section are the basis for the following caching-determination procedure.
In this section, the core problem---how to extract or predict the topics from the existing tweets based on machine learning approaches, is taken into consideration. This section is separated into three approaches of solving the problem. In the first segment, we illustrate how the LDA approach is exploited to extract latent topics from the existing tweets. In the second segment, we demonstrate how to chronologically predict future tweets based on LSTM. In the last segment, we display how to exploit our new embedding method to chronologically predict future tweets associated with geographical locations based on LSTM.

\subsection{Extracting Latent Events from Tweets}
The introduction section has demonstrated that the twitter events are associated strongly with the public preference in prior works, in which approach are we able to extract the events with existing twitter text is a crucial term towards realising our proposed wireless caching framework. In this section, we demonstrate how to exploit LDA method to obtain satisfying events extraction.
\subsubsection{The LDA model}
\begin{figure}[htbp]
   \begin{center}
        \includegraphics[width=2.5in]{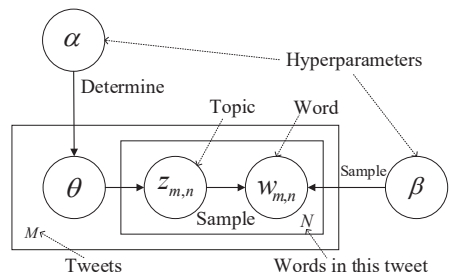}
        \caption{The structure of LDA model for latent events forecasting.}
        \label{LDA_stru}
    \end{center}
\end{figure}

LDA is an unsupervised Beyesian probabilistic model with the objective to identify the probable topics from the documents~\cite{seven_blei2003latent}. The proposed LDA model for latent events forecasting is illustrated in Fig. \ref{LDA_stru}. As in Fig.~\ref{LDA_stru}, the idea is to determine the corresponding topic $z_{m,n}$ of the $n^{th}$ word in $m^{th}$ document based on parameter $\alpha$ while to determine the probability $w_{m,n}$ of each word under the given parameter $\beta$ and a given topic in order to obtain every word in this given topic.
Each document is assigned a distribution to the latent topics. Through setting the $K$ latent topics, the words are associated with the topics as $P(topic|document)$ and $P(topic|term)$ \cite{eight_fang2016topics}.

The objective is to obtain the probabilities that each word belongs to different latent events, and therefore, we allocate events by listing the words of the highest probability in each latent event, illustrated in Eq. \eqref{lda},
\begin{align}\label{lda}
P(Z\vert{ W,\alpha,\beta}),
\end{align}
where $Z$ is the event, $W$ stands for the document, and $\alpha$, $\beta$ are parameters.

According to Fig.\ref{LDA_stru} and its approach originated from \cite{seven_blei2003latent}, the probability equation is given by Eq. \eqref{theta_equ},
\begin{align}\label{theta_equ}
p(\theta |\alpha ) = \frac{{\Gamma (\sum\nolimits_i {{\alpha _i}} )}}{{\prod\nolimits_i {\Gamma ({\alpha _i})} }}(\prod\nolimits_i {\theta _i^{{\alpha _i} - 1}} ),
\end{align}
where $\theta$ is the event multinomial distribution parameter, $\alpha$ is a k-dimensional vector of the Dirichlet districution. $\theta$ complies to the Dirichlet distribution of parameter $\alpha$.

With given parameter $\alpha$ and $\beta$, the joint probability distribution is able to be demonstrated in Eq. \eqref{joint_equ},
\begin{align}\label{joint_equ}
p(\theta ,z,w|\alpha ,\beta ) = p(\theta |\alpha )\prod\nolimits_{n = 1}^N {p({z_n}|\theta ) \cdot p({w_n}|{z_n},\beta )},
\end{align}
where $z$, $w$ stand for event set and document set, $N$ stands for the $N$ words in document $w$. As $\theta$, $z$ are latent variables, we adopt Gibbs Sampling to marginalize them.

\subsubsection{Model Learning}
The learning procedure of the LDA model is based on the Gibbs Sampling. The learning procedures of the LDA model are demonstrated in \textbf{Algorithm~\ref{LDA}}. Here, we set the quantity of iterations to 100, the size of the latent topics to 20, and words per topic to 7.

\begin{algorithm}[!t]
\caption{LDA Learning for Latent Events Forecasting}
\label{LDA}
\begin{algorithmic}[1]
\STATE For every word $w$ in each document, assign the word to a random latent topic $z$;
\REPEAT
    \STATE Scan the corpus, for every word $w$ in the corpus, do Gibbs Sampling, and obtain its latest latent topic;
    \STATE update the corpus;
\UNTIL{The Gibbs Sampling Converges OR Reach the maximum iteration number}
\STATE Gather the topic-word ($W$-$Z$) Co-occurrence Matrix, which is the result of LDA model;
\STATE Obtain the topic distribution of the model $\vec{\theta}_{new}$;

\end{algorithmic}
\end{algorithm}

\subsection{Forecast Events based on Tweets}
\begin{figure}[htbp]
   \begin{center}
        \includegraphics[height=2.5in]{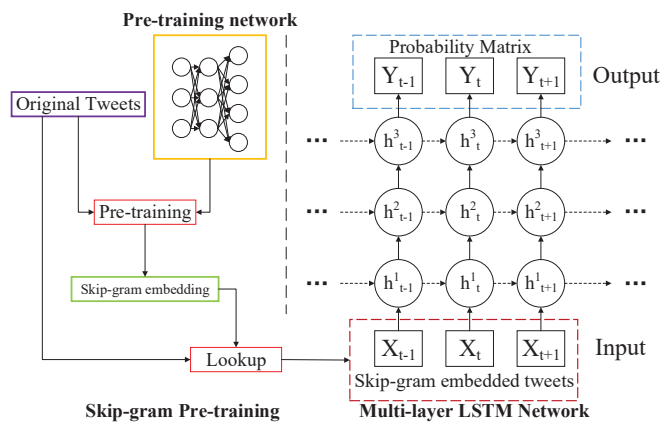}
        \caption{LSTM model for content popularity forecasting}
        \label{lstm_structure}
    \end{center}
\end{figure}

LSTM is compatible for chronological data forecasting. Compared with conventional RNN, the LSTM model has the advantage in memorizing the long-term memory~\cite{ten_cho2014learning}. The structure of the proposed LSTM model is illustrated in Fig.~\ref{lstm_structure}. In Fig.~\ref{lstm_structure}, LSTM cells are basic units of the LSTM model. The two arrows surrounding the cell $Cell_{t}$ are the vector transfer of the long-term memory $c_{t-1}$ from the former cell $Cell_{t-1}$ and the short-term memory $h_{t-1}$ from the former cell $Cell_{t-1}$. Inside the cell are neural network Layers, including tanh layers and sigmoid layers. $X_{t}$ and $h_{t}$ are the input and output at the time $t$. The algorithm flow of training the LSTM model is listed in \textbf{Algorithm~\ref{LSTM_Algo}}

\begin{algorithm}[!t]
\caption{LSTM for Content Popularity Forecasting}
\label{LSTM_Algo}
\begin{algorithmic}[1]
\STATE Initial the model layers and hidden sizes with given parameters;
\STATE Feed the first batch of input $X_{1}$, calculate the memory $h_{1}$ and Cell State $C_{1}$;
\REPEAT
    \STATE For time stamp $t$, the forget gate layer $\sigma$ determine what parts of the cell state $C_{t-1}$ are preserved;
    \STATE For time stamp $t$, the input gate layer $\sigma$ determines which value between $h_{t-1}$ and $x_{t}$ is updated. And derive the cell state of this time stamp $C_{t}$, memory $h_{t}$ and candidate value $\widetilde{C}_{t}$ through $tanh$ layers and operators;
    \STATE Invoke a sigmoid layer (output gate layer) to determine what parts of the cell state are the output and update the cell state;
    \STATE Evaluate perplexity through calculating cross entropy between the forecasting and $X_{t+1}$;
    \STATE Invoke the gradient descent optimizer with a given learning rate to optimize the model;
\UNTIL{Reach the maximum epoch number}
\STATE Save the model;
\end{algorithmic}
\end{algorithm}

The skip-gram embedding approach is commonly employed in natural language processing (NLP). The idea is to feed the model with the words of adjacent positions and therefore associate the words with the context. For example, while skip is 1 and a batch is n+1 words, we firstly feed the model with the range of $i^{th}$ to ${(i+n)}^{th}$ words of the text, and the second step is to feed the ${(i+1)}^{th}$ to ${(i+n+1)}^{th}$ words~\cite{eleven_zaremba2014recurrent}.

\subsubsection{Model Training}
To train the LSTM model, we exploited the Twitter dataset from 27th January 2018 to 26th February 2018 (30 days) as the training dataset, and tweets of 27th February 2018 as the testing dataset. Here, the dictionary is limited to be 60000 words. To fulfill the prerequisites, each tweet is mapped into a vector based on the indexes of each word in the dictionary. The vectors are fed into the network based on their chronological order. The parameters of the networks are detailed in Table~\ref{skip para}. We use a gradient descent optimizer in TensorFlow to automatically adjust the learning rate.

\begin{table}[htbp]
\caption{Parameters of LSTM models}
\begin{center}
\begin{tabular}{|c|c|c|}
\hline

\cline{2-2}
\textbf{\textit{Parameter}}& \textbf{\textit{Medium}} &\textbf{\textit{Large}}\\
\hline
Initial scale of weights & 0.04 &0.05\\
Initial learning rate & 0.1&0.2\\
maximum permissible norm of the gradient & 5&10\\
number of LSTM layers & 2&3\\
number of unrolled steps of LSTM & 20&50\\
hidden size & 650&1500\\
max epoch & 65&55\\
learning rate decline epoch &25&30\\
learning rate decline rate &0.8&$2/3$\\
\hline
\multicolumn{2}{l}{}
\end{tabular}
\label{skip para}
\end{center}
\end{table}

\subsubsection{Forecasting}
The model forecasts the terms of the events through a softmax (normalized exponential) layer, which is commonly used loss function for multi-class classification \cite{bishop2006pattern}. The events are sorted according to the predicted probabilities and the events with higher probabilities are cached in our framework.

\subsection{Predicting Events based on Tweets and Geographical Locations}
The anticipation of this model is to combine the geographic information with the Twitter text. To achieve the objective, we append the tokens indicating the latitude ranges and longitude ranges to the tweet vectors. By feeding the LSTM model with the embedded vectors including the tokens, the text are therefore able to be associated with the geographic information.

\subsubsection{Model Training}

\begin{figure}[htbp]
   \begin{center}
        \includegraphics[width=3.4in]{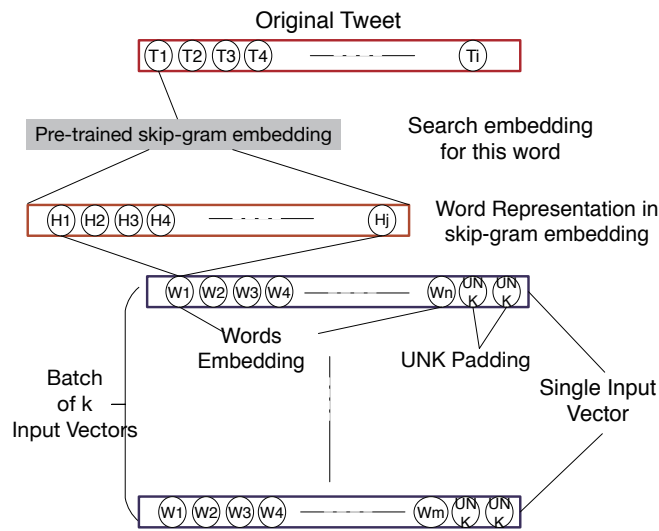}
        \caption{LSTM inputs based on skip-gram embedding.}
        \label{cbow}
    \end{center}
\end{figure}

In this new model, we apply BoW embedding to map each tweet into a vector. Regardless of the order of the components, we merely assume the terms are ingredients of the topic of the tweet. The skip-gram approach is invoked to feed the vector of tweets into the model. The process is illustrated in Fig. \ref{cbow}. Fig. \ref{cbow} presents the inputs of the proposed LSTM for latent events forecasting. In Fig. \ref{cbow}, each single tweet is mapped into a vector based on its content and the geographic information of its BS. Thus, for the first $n$ elements ($W_{1}$ to $W_{n}$) in single tweet vector, they are the numerical representation of the words in this tweet based on the index in the above dictionary.

After the data cleaning procedure mentioned in data preparation section (removing the meaningless parts in tweet text), most tweets are able to be restrained within 20 words (approximately 97\%). While larger vector leads to lower accuracy, we constrain the vector size to 20 elements. For those less than 20 words, UNK tokens are filled into the vector. For tweets longer than 20 words, we truncate the first 20 words where the tags and most words are preserved. Afterwards, the last two elements in the tweet vector represent the latitude range and the longitude range of the tweet. The Latitude token ranges from 81112 to 81114 while the Longitude token ranges from 91112 to 91114 (to avoid the collision with the indexes of the dictionary). Afterwards, every $m$ tweet vectors are arranged into the single input vector based on their chronological order. Between the single vectors, we invoke the skip-gram approach to maintain the chronological relations among the tweets as illustrated in Fig. \ref{cbow}. The second input vector includes $V_{n+1}$ to $V_{n+m+1}$ tweet vectors ($V_{n}$) along with the first single vector includes $V_{n}$ to $V_{n+m}$ tweet vectors. Every $k$ single input vectors are deemed as a batch. The LSTM model itself is the basic LSTM neural network based on TensorFlow. Here, we evaluate the dataset through the LSTM models of two scales. The basic parameters of the models are listed in Table \ref{cbow para}.

\begin{table}[htbp]
\caption{Parameters of LSTM models}
\begin{center}
\begin{tabular}{|c|c|c|}
\hline

\cline{2-2}
\textbf{\textit{Parameter}}& \textbf{\textit{Medium}} & \textbf{\textit{Large}}\\
\hline
Initial scale of weights & 0.04 & 0.05\\
Initial learning rate & 0.1 & 0.2\\
maximum permissible norm of the gradient & 5 & 10\\
number of LSTM layers & 2 & 3\\
number of unrolled steps of LSTM & 50 &100\\
hidden size & 650& 1000\\
max epoch & 45 & 55\\
batch size & 20 & 20\\
learning rate decline epoch &25&30\\
learning rate decline rate &0.8&$2/3$\\

\hline
\multicolumn{2}{l}{}
\end{tabular}
\label{cbow para}
\end{center}
\end{table}

\subsubsection{Predicting Geo-information and tweet content}
The softmax layer is utilized to predict the topic according to the descending order of the probability of the words. The prediction only contains the words in the dictionary. To predict the geographic information that in which of the 9 areas the tweet is posted, we sum up the probability of each geographic elements in the single tweet vectors. As the output of the softmax layer is a $(k\cdot{m\cdot{(n+2)}})\times{vocabulary\_size}$ matrix and the $k,m,n$ are illustrated in Fig. \ref{cbow}. The vocabulary size is set to be 92000 to include all the indexes including both dictionary indexes and geographic indexes. Then for every $n+2$ rows (this represents a single tweet), we separately sum up the $81112^{th}$ , $81113^{th}$ and $81114^{th}$ columns of these $n+2$ rows. The greatest value among the three indicates that the tweet belongs to the corresponding latitude area. The determination of the longitude is of the same approach.

\section{Practical Experiments for Latent Events Forecasting}
In this section, practical experiments are demonstrated to evaluate the three models utilizing the collected data from Twitter. The three models generate different types of outcomes: 1) The LDA model predicts latent events. 2) The LSTM model with skip-gram embedding forecasts related words of events. 3) The LSTM model with skip-gram-geo-aware embedding predicts related words of the event and its location.

To demonstrate the performance of the proposed models, we adopt the perplexity value as the key performance indicator, because perplexity is an well accepted approach to evaluate natural language processing (NLP) models. The definition of perplexity is illustrated in Eq.~ \eqref{perp}:

\begin{align}\label{perp}
{2^{H(p,q)}} = {2^{ - \frac{1}{N}\sum\limits_{i = 1}^N {{{\log }_b}q({x_i})} }},
\end{align}
where $p$ is the unknown distribution of the test dataset, $x_{1}$, $x_{2}$, $x_{3}$,...,$x_{N}$ are subsets of test dataset. $q$ stands for the model that we want to evaluate. Perplexity evaluates the similarity between the prediction and the ground truth. Since perplexity maintains a reciprocal relationship with the Log-likelihood measures, a language model with lower perplexity achieves better performance during application. Therefore, in this section, we apply perplexity to demonstrate the accuracy of the proposed models and the testing datasets are clarified in each sub-section. Furthermore, the unified standard---complexity is also capable of generalizing the results since the standard is based on numerical results.

\subsection{LDA model}
\subsubsection{Dataset division}
\begin{figure}[htbp]
   \begin{center}
        \includegraphics[width=3.5in]{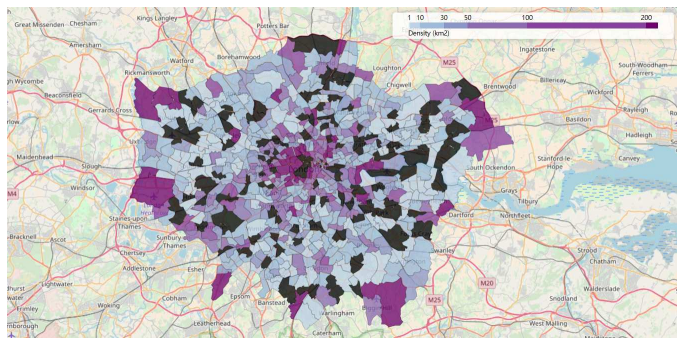}
        \caption{The density distribution of the involved BSs in London (Area marked with deeper purple contains more BSs). We observe that the areas of City of London and Inner London are equipped with denser BSs compared with areas of Outer London.}
        \label{heatmap}
    \end{center}
\end{figure}

To demonstrate the distinct preference of topics within different regions, the dataset has been divided into 9 smaller datasets. The separation is based on the BS coordinates which is illustrated by Fig.\ref{heatmap}. According to Fig.~\ref{heatmap}, areas of City of London and Inner London are equipped with denser BSs compared with areas of Outer London.

\subsubsection{Literal results}
To demonstrate the variation of different extracted topics in different regions and considering the page limit, we display the results in Location 5 and Location 4 for examples. The reason for selecting the two locations is that the topics in the two regions are more diversified compared to other regions.The literal results of the LDA model in Location 5 are demonstrated below in Table \ref{result_4_5}. The results from the location 5 are able to be interpreted in the following approach. The first topic relates to employment advertising tweets with the tag of ``\#careerarc" .The $2^{nd}$ and $3^{rd}$ topics associate with the equality of the ``LGBT" group and the weather.
\begin{table}[htbp]
\caption{Sample results in different regions generated from the proposed LDA model}
\begin{center}
\begin{tabular}{|c|c|}
\hline

\cline{2-2}
\textbf{\textit{Region}}& \textbf{\textit{Topics}}\\
\hline
1& park hyde loving fc tottenham gunners children\\
2& stadium teaching wembley fitness art centre 14\\
3& rain mm mph wind hpa thames fine\\
4& london greater heathrowairport station hounslow new city\\
5& job hiring england careerarc london latest work\\
6& park studio team join love old o2\\
7& uk egaylity 12 gay city stigmabase wembley\\
8& essex palace art teaching gallery north en\\
9& london greater bridge station house free unitedkingdom\\

\hline

\multicolumn{2}{l}{}
\end{tabular}
\label{result_4_5}
\end{center}
\end{table}
According to the results in Table \ref{result_4_5}, the first and the third topic are possibly originated from the tweets composed by tourists as they are discussing the particular locations like the Heathrow Airport and Hounslow. The second topic is related to the Valentine's Day, however as there is also a music festival called ``Kaleidoscope Festival" during the same time period, the keyword ``kaleidoscope" is also included here.

\begin{figure}[htbp]
   \begin{center}
        \includegraphics[width=3.4in]{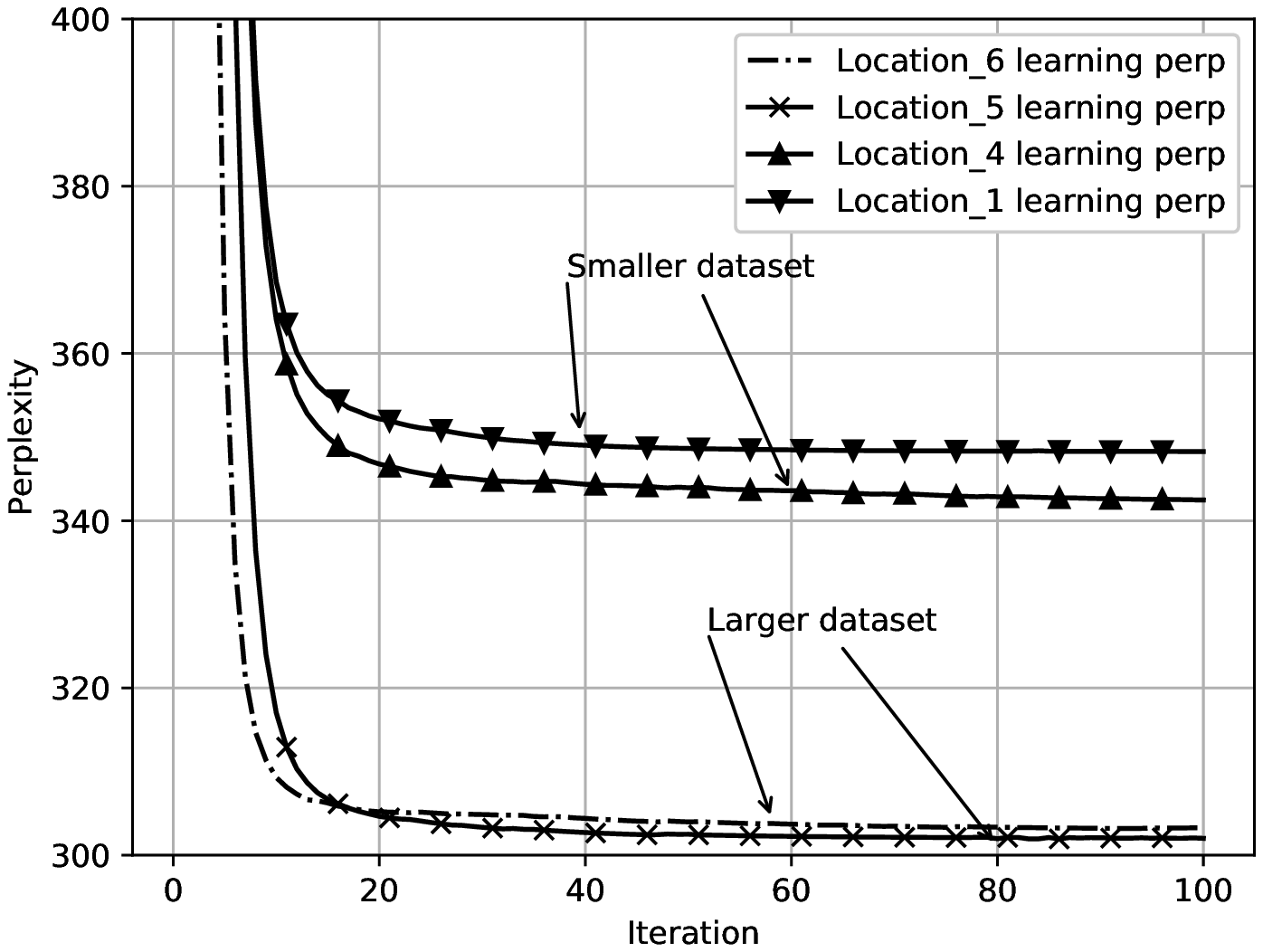}
        \caption{The learning perplexity of the proposed LDA model in Location 1 ,Location 4 (smaller datasets), Location 5 and Location 6 (larger datasets).}
        \label{LDA_perp}
    \end{center}
\end{figure}

The complexity tendencies of the LDA model under different datasets are illustrated in Fig.\ref{LDA_perp}. To illustrate the different tendencies of perplexity under datasets of different sizes, we select two larger datasets and two smaller datasets based on their regions. The datasets of location 6, location 5 are larger datasets while the rest two datasets are smaller ones. The graph demonstrate the results in two aspects. The decreasing trends demonstrate that the LDA model is able to converge and the larger datasets enable the model to achieve more accrate performance (lower complexity).

\begin{figure}[htbp]
   \begin{center}
        \includegraphics[width=3.4in]{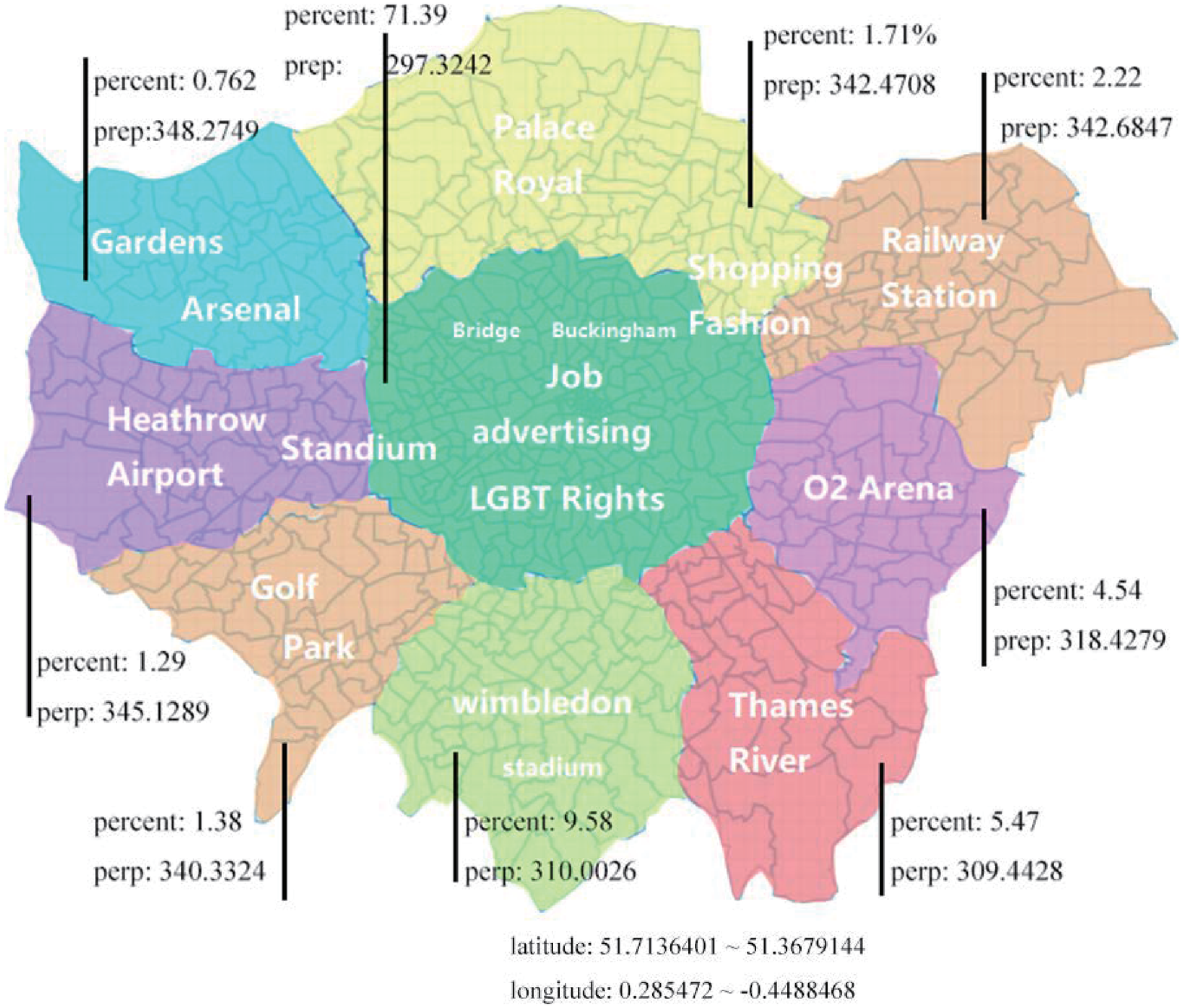}
        \caption{Keywords distribution of the predicted events in London generated from the proposed LDA model.}
        \label{colored}
    \end{center}
\end{figure}

Here are we demonstrate some keywords of the events from LDA model as illustrated in Fig. \ref{colored}. From the graph, tweets in the 9 areas are primarily related to the landmark places or district functions (sightseeing, sports), which assists to allocate different characteristics among the districts in London.

Analysis of the keywords is capable of determining the text-related caching contents based on the public preference. Since we have discovered that the keywords from tweets in different regions are associated closely with the indigenous landmarks, the text-related contents are able to be determined from them. For instance, according to the Fig. \ref{colored}, in the south part, the keywords ``wimbledon" and ``stadium" are illustrated. Different from the conventional counting methods, our proposed machine learning approach is capable of demonstrate the concurrency among the words. Therefore, based on the two keywords mentioned above, the text-related contents, such as videos for the matches held in the stadium, are supposed to be cached at nearby base stations. Similarly, for the base stations located near the keyword ``Arsenal", the related videos of the recent football games should be cached. Different from the situations mentioned above, regarding to the palaces in the central part of London, the pictures and navigation information (including coordinates and public transport information) should be cached since there are several sightseeing spots including the famous Buckingham Palace.

\subsection{LSTM model with skip-gram embedding}
\subsubsection{Datasets}
The 9 training datasets and 9 testing datasets are separated based on regions of BS. The training datasets range from $27^{th}$ January to $26^{th}$ February. The testing datasets are tweets on the $27^{th}$ February.
\subsubsection{Forecasting results}
For the similar reason with the literal results in LDA model, we demonstrate the results from Location 4 and Location 5. The top events of the location 5 and location 4 are illustrated in the Table \ref{skip_result}. Same as LDA model, as the perplexity trends converges during the testing, the generalization and the accuracy of the model are able to be demonstrated. The interpretations are based on the keywords in the topics. With the keyword ``job", ``hiring", ``\#CEBCareers", the first topic is able to be considered relating to the job advertising while containing noise from irrelevant topics. The second topic involves a couple of landmarks, like ``Belgravia", ``Sleaford" and ``Langham" along with meteorological terms like ``UV", ``Rain", ``wind". Therefore the topic is considered discussing the weather corresponding to the landmarks. To generalize the results, numerical results are able to provide evidence. Since the perplexity of this model tends to converge and the final complexity values are satisfactory, the accuracy of LDA models are demonstrated.

\begin{table}[htbp]
\caption{Sample result events from the proposed LSTM model with skip-gram embedding}
\begin{center}
\begin{tabular}{|c|c|}
\hline

\cline{2-2}
\textbf{\textit{Location}}& \textbf{\textit{Topic}}\\
\hline
 \multirow{4}{*}{Location 5}& \#job \#Hiring \#CEBCareers May UK \\& marriage Northern \#BusinessMgmt urged\\& Ireland take equal forward Manager \\& \#London anyone recommend Pourtsmouth \#makeupartist May\\

\hline
 \multirow{4}{*}{Location 4}& \#London United Kingdom The Belgravia\\ & Click \#ProductMgmt damn today mph fine\\& UV Sleaford Langham England Rain We're\\& \#london \#giftshop 2018\\

\hline
\multicolumn{2}{l}{}
\end{tabular}
\label{skip_result}
\end{center}
\end{table}

\begin{figure}[htbp]
   \begin{center}
        \includegraphics[width=3in]{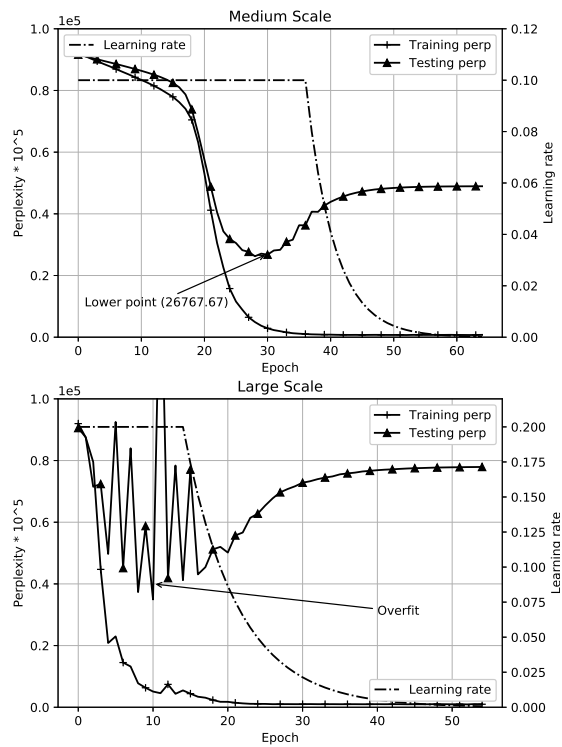}
        \caption{The training and testing perplexity under medium scale and large scale of the proposed LSTM networks.}
        \label{loca5_lstm}
    \end{center}
\end{figure}

While invoking the model on the location 5 dataset with networks of different scales, the variation of the training and testing perplexity is illustrated in Fig. \ref{loca5_lstm}.

\subsection{LSTM model with skip-gram-geo-aware embedding}
\subsubsection{Datasets}
The training dataset contains all tweets from $27^{th}$ January to $26^{th}$ February. The testing dataset is the tweets on the $27^{th}$ February.
\subsubsection{Forecasting results}
\begin{figure}[htbp]
   \begin{center}
        \includegraphics[width=3in]{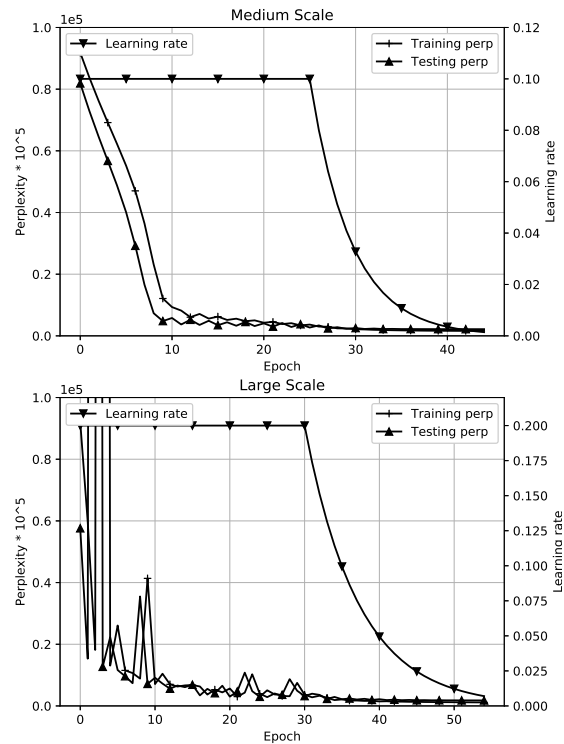}
        \caption{The training and testing perplexity under medium scale and large scale of the proposed LSTM networks (Geo-aware).}
        \label{cbow_perp}
    \end{center}
\end{figure}

The training and testing perplexity tendency are illustrated in the Fig. \ref{cbow_perp}. The text prediction examples are illustrated in Table \ref{cbow_result}. The topic is largely similar to the results of invoking the LDA model to the Location 5 dataset. As the job advertising tweets constantly contain a URL directing to their web site, the keywords ``apply" and ``click" are involved in the prediction. The prediction is more unified and of less noise compared to the results of the skip-gram embedding.

\begin{table}[htbp]
\caption{Result Topic and Location examples}
\begin{center}
\begin{tabular}{|c|c|}
\hline

\cline{2-2}
\textbf{\textit{Location Forecasting}}& \textbf{\textit{Topic}}\\
\hline
 \multirow{3}{*}{Location 5} & \#London \#CareerArc United \#Hiring England \\ &The \#job UK Kingdom work See latest Greater \\ &I'm opening Can apply Click\\

\hline
\multicolumn{2}{l}{}
\end{tabular}
\label{cbow_result}
\end{center}
\end{table}

\subsection{Model Analysis}
In this section, the properties of the three models are analyzed. The performances of models under different circumstances are demonstrated. Based on the graphs mentioned in the above section, we discuss the different aspects of performance illustrated.

\subsubsection{LDA model with different datasets}
The perplexity applied in the LDA model is able to be regarded as a standard to evaluate the cohesiveness between the extracted events and the ground truth---the existing tweet text. Since complexity is related tightly to the Log-likelihood, less complexity leads to more cohesiveness between the extracted topics and the twitter text and more accuracy of the results. The convergence of the complexity is also able to be deemed the convergency of the model. With different scales of documents illustrated in Fig.\ref{LDA_perp}, larger documents (Location 5 / 6) tend to converge more rapidly. Furthermore, the overall complexity of larger documents tend to be lower. Since more tweets have involved and during each iteration, for definitive $K$ latent topics and a larger corpus, the topic-word co-occurrence matrix is more reasonable. The final perplexity of the LDA model is satisfactory (about 300 under large dataset). Since the literal results demonstrated in Table \ref{result_4_5} contain correlative key words, we can interpret the topics expediently. Regarding the advantages above, we demonstrate that the LDA model is quite feasible of predicting Twitter topics.

\subsubsection{Different scales of LSTM with skip-gram embedding}
Different from the LDA model, less complexity of the model when applied on the testing datasets leads to less difference between the prediction and the ground truth---the testing datasets (future tweets). The convergence of the complexity is also able to deemed the convergency of the model.
Under larger scale of network and learning rate, the tendency in Fig.\ref{loca5_lstm} tends to converge more rapidly while the testing perplexity fluctuates obviously due to overfitting and the characteristics of the gradient descend optimization. As mentioned above, complexity is applied to evaluate the accuracy of the models in this paper. According to the result, the smaller network and learning rate result in better overall complexity. The final testing complexity of the medium-scale network is 54738.23 while that of the large-scale network is 78312.39. Due to the polytropic combination of oral words, the perplexity of the Twitter documents are not restrained as that of the formal dataset like PTB (about 70 perplexity in large-scale LSTM model \cite{eleven_zaremba2014recurrent}).

\subsubsection{Different scales of LSTM with skip-gram-Geo-aware embedding}
When the new model converges, the topics of prediction are largely the same due to the characteristics of the LSTM model;therefore, the model itself is only capable of predicting one tweet in a specific area. The variations of the complexity under different scales of networks are demonstrated in Fig.\ref{cbow_perp}. The larger network with greater initial learning rate tends to converge more rapidly while fluctuating obviously. After the training and testing process, the larger network with greater depth of network structure concludes superior results on the overall testing perplexity. The outcome demonstrates that the larger network tends to generate more monolithic outcome results with the testing dataset. The final testing perplexity is 1253.75(Larger model) / 1752.83 (Medium model).

Based on the events prediction and perplexity results of the three models, the LDA model achieves satisfactory prediction as well as multiple topics prediction within one particular area. While the basic LSTM model with skip-gram embedding is of relatively high perplexity. The novel LSTM model with skip-gram-Geo-aware embedding realizes comparatively lower perplexity and the geographic information prediction. However, due to the characteristics of the neural network when converges, future efforts are required to enable the model to predict multiple topics.

\section{Associating Events Forecasting with Caching}

Three machine learning models are proposed in above sections to extract latent events from the existing tweets and predict future topics based on existing tweets. The perplexity results confirm the effectiveness of the proposed models. Since we have proposed valid approaches to solve the core problems mentioned above, we need then to evaluate the model in the context of wireless caching. In this section, we associate the events prediction with the wireless caching. To exploit different advantages of different machine learning models, we propose the following approach to determine the text-related caching contents. While the LSTM-based models mentioned above are capable of predicting the future twitter text based on the chronological inputs of existing tweets, the LDA model is capable of extracting the latent topics from the predicting twitter text. Afterwards, the text-related contents---the images and videos, which are capable of generating remarkable impact on the network backhaul loads, are cached at the BS. Since the accuracy of the framework is directly related to the accuracy of the LDA model, we mainly discuss the accuracy of LDA topic-extracting model as the machine learning approach (ML) in this section.

The hit rates is applied to evaluate the performance of the proposed framework. Since determining text-related contents has no previous algorithms, we take conventional algorithms (LFU, LRU) for comparison. The LFU approach is applied to extract the most frequently used keywords from the existing tweet text while the LRU obtains the most recent keywords from the existing tweet text. The keywords extracted through the two conventional methods are applied to match actual media files. As for evaluation criteria, we proposed the "hit portion", which is the portion of utilized topics among all the extracted topics. Regarding the specific caching contents evaluation in this section, we demonstrate the difficulties of determining the caching contents of social media contents as well as propose our criterion to evaluate the model. Therefore, the coherency between the topics and caching contents is better demonstrated. Here, the testing datasets are the tweets on 27th February while the training datasets are the tweets ranged from 27th January to 26th February.

\subsection{TAC strategy}
In this section, we demonstrate the strategy for caching actual media files (images and videos) regarding the twitter topics.

\subsubsection{Caching objects}
Since we have demonstrated the structure of our framework at the previous section and the approaches to obtain topics prediction have been demonstrated, we introduce the caching strategy to associate the topics with actual caching contents (images, videos). As the tweet text has been cleaned in the dataset preparation procedure, the words in tweet text are seldom meaningless and each of the words represents a symbol of identification. The core idea is to decide which tweet is valuable to be cached through predicted topics. To judge whether a topic is associated with a tweet, we define when there are 3 words the same between a topic and text of a tweet, the media files associated with this tweet are worthwhile to cache. After determining the tweet has the value to be cached---hit by the topics, the media files are cached through the URLs retrieved from the JavaScript Object Notation (JSON) data structure, which is obtained from Twitter API.

\subsubsection{Prior list}
After gathering the actual media files and extracting topics from existing tweet text, the relationship between the two parts is established. The popularity of the topics and actual caching objects is capable of being definitely settled through HTTP requests from users.
Since there are generally large number of topics contained in tweet text and topics themselves are of popularity of different scale, the ``Prior List" (PL) is applied to filter the topics of the most popularity to maximum the caching efficiency with certain number of topics and caching space for actual media files. Similarly, a PL of caching objects is a ranking list contains actual caching objects related to a specific topic obtained through Twitter API.
The rankings of caching objects are based on their frequency of usage, namely their popularity. Regarding the existing caching strategy, which applies popularity of caching objects\cite{twenty-seven_6883600}, the efficiency of the caching algorithm increases. With the aim of precisely determine what to cache, we create a PL for each topic to preserve the media files related to it. The objective of the PL is to associate the extracted topics with the popularity of the caching objects, which improves the efficiency of the caching framework.

As the BS is capable of monitoring the wireless traffic through HTTP requests / responses, PLs related to different topics are varying dynamically after deploying our caching framework at the BS. When the topics PL has reached maximum amount of topics and a new topic emerges, the popularity of the new topic is compared with those of topics in the topics PL. When the popularity of the new-emerged topic exceed any from those of the topics in PL, the PL is updated. Moreover, the caching space is updated.

\subsection{Evaluating the extracted topics}
In this section, the performance of the ML approach as well as the conventional LFU, LRU approaches is evaluated through four different numerical results---tweet hit rate, tweet hit portion, cache portion and hit cache portion. The LFU and LRU approaches extract the keywords from the tweet text rather than the topics like the ML approach.
The aim of this section is to compare the performance of the models as well as evaluate the properties of the ML approach under different circumstances.

\subsubsection{Tweet hit rate}
\begin{figure}[htbp]
   \begin{center}
        \includegraphics[width=3.4in]{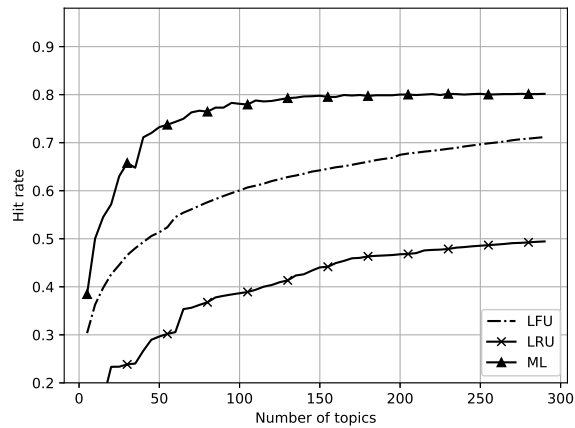}
        \caption{The hit rate of topics of different approaches.}
        \label{Total_hit_rate}
    \end{center}
\end{figure}

In this part, we demonstrate hit rates of tweets in our model. The aim for considering tweet hit rates is to illustrate to what extent a model can associate the topics with the tweet text. In Fig.\ref{Total_hit_rate}, the hit rates of three approches are demonstrated---LFU, LRU and the Machine Learning approach (ML). The X-axis of the graph is the amount of topics. By fixing the number of topics extracted from the tweet text, we are therefore capable of discussing the performances under different caching space since more topics lead to more media files to be cached. Rather than merely extracting keywords from the corpus like LFU and LRU, ML approach is actually extracting topics, i.e. the cohesive keywords from the corpus, which assist the approach achieve high coherency between the topics and the tweet text. Therefore, the ML approach achieves the satisfactory results towards extracting topics of tweet text from the corpus.

With the aim of further developing the properties of the machine learning approach, we discuss the performance of the method under different circumstances.
To demonstrate the different trends under datasets of different sizes, we select 4 datasets from different regions to illustrate the trends. The selection of the datasets are based on their sizes (number of tweets) with the aim to invest the model property in different dataset ranges. Among the 4 testing datasets exploited to evaluate the model, the sizes of training datasets are location 5 $>$ location 8 $>$ location 9 $>$ location 6.
\begin{figure}[htbp]
   \begin{center}
        \includegraphics[width=3.4in]{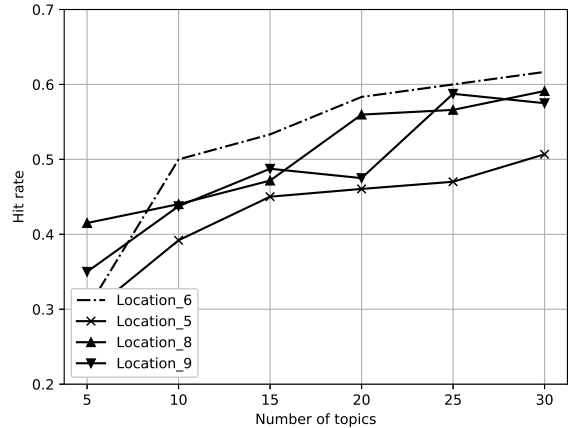}
        \caption{The hit rate of topics in different regions.}
        \label{LDA_hit_rate}
    \end{center}
\end{figure}

In Fig. \ref{LDA_hit_rate}, the trends of hit rates can be described as below. 1) The hit rates of the model on all testing datasets increases when the numbers of topics increases. 2) The overall hit rates vary little among the different sizesof datasets, which means that the little variance among the testing datasets does not influence the overall accuracy. 3) The highest hit rates are achieved when there are maximum number of topics and the highest hit rates vary from 50\% to 65\% under different testing datasets. As a conclusion, the wireless caching framework we propose is capable of generating relatively satisfactory results.

\subsubsection{Tweet hit portion}
To display the utilization ratio of the topics based on the three approaches mentioned above---LFU, LRU and Machine Learning approach (ML). The tweet hit portion is the utilization ratio of the topics, namely the topics hit by the tweet text of the future. The aim of introducing this property is also to demonstrate the effectiveness and accuracy of the abilities to extract topical information from the corpus. We define the tweet hit portion to be the portion of topics hit by the testing tweet text dataset. The aim of this criterion is to illustrate the accuracy of the topics generated from different approaches. Besides, the performance of the ML approach under different datasets is discussed.

\begin{figure}[htbp]
   \begin{center}
        \includegraphics[width=3.4in]{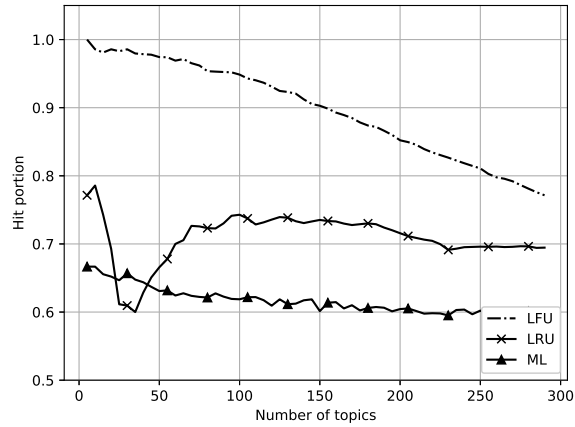}
        \caption{The hit portion of topics of different approaches.}
        \label{Total_hit_port}
    \end{center}
\end{figure}

In Fig.\ref{Total_hit_port}, the machine learning approach (ML) achieves the lowest hit portion rate while the LFU and LRU achieve higher hit portion. Therefore, the contradiction is that the machine learning approach achieves higher tweet hit rates as well as the lower hit portion under the same topics numbers compared to the other two methods. The explanation to this contradiction is that while the LFU and LRU focus on more monotonic topics---The recent ones and the hot ones, the topics from these two approaches are more unified compared to the ML approach. While the ML approach generates a more heterogeneous topics prediction, the hit portion is lower compared to LFU and LRU.

\begin{figure}[htbp]
   \begin{center}
        \includegraphics[width=3.4in]{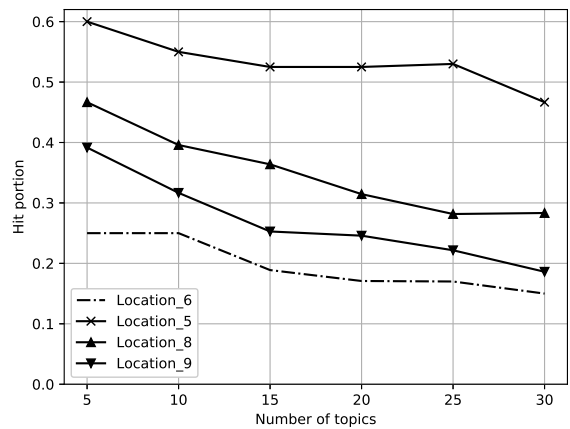}
        \caption{The hit portion of topics in different regions.}
        \label{LDA_hit_port}
    \end{center}
\end{figure}

To invest the ML approach properties under datasets of different sizes, we choose the 4 datasets from different regions. The selection of the datasets are based on the familiar reason as that of the tweet hit rates section. In the Fig. \ref{LDA_hit_port}, the trends of hit portions are demonstrated. The results are able to be explained from the following aspects. 1) The utilization ratios of the caching topics decrease along with the increasing number of extracted topics. 2) The over all trends of utilization ratios vary remarkably among different sizes of testing datasets. Moreover, larger datasets (location 5,8 datasets) are able to achieve higher utilization ratios. 3) The highest utilization ratios are achieved when there are least extracted topics and the maximum utilization ratio is approximately 60\% with the largest testing dataset of location 5.

\subsection{Problem of evaluation and proposed solution}
During the collection of the datasets, we noticed an obstacle of evaluating the effectiveness and accuracy of our model. The tweets with geo-tags are largely the original ones, which are created by user themselves rather than retweet. This situation leads to the conclusion that the overlap of media files between the two different days are particularly little. Different from conventional video sites, the contents of the social platforms are largely published by the public, which makes the determination of caching contents nontrivial.
However, since the traffic of social media platforms is capable of being formulated into two aspects---viewing (downloading) and posting (uploading), we propose a unified approach to evaluate the model. Users prefer to not only go through the contents (text, images, videos) that they favor, but also to post their own contents associated to the topic, such as a video related to a popular kind of pet in that region. Therefore, when media contents of the next day are similar to the media contents at the present day, the media contents cached today are highly possible to be viewed by the users.
Regarding to the reasoning above, we propose the two evaluation criteria, namely ``cache portion" and ``hit cache portion" to evaluate the models.

In this section, the performance of the models is evaluated through solid text-related caching contents(images, videos). To process the evaluation, the numerical results are presented from two different aspects.

\subsubsection{Cache portion}
First part is the cache portion. Cache portion is the portion of media files (text-related caching contents) which are cached after the caching-content determination procedure. This criterion represents how much of the formal caching contents is cached based on the topics, which relates tightly to the size of occupied caching space at BS. With higher caching portion, the requirements for caching space are higher---more contents are cached. Here, the portion is the size of files that is cached divided by the total size of media files in the training dataset.

\begin{figure}[htbp]
   \begin{center}
        \includegraphics[width=3.4in]{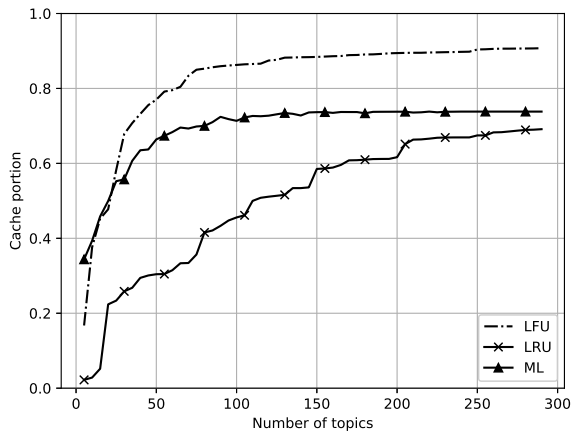}
        \caption{The caching portion of the training dataset (different approaches).}
        \label{Total_size_train}
    \end{center}
\end{figure}
In Fig.\ref{Total_size_train}, the cache portion of the 3 approaches---LFU, LRU and Machine Learning approach (ML) are illustrated. In the figure, LFU maintains the highest cache portion up to about 90\%, which leads to the conclusion that under our caching strategy, 90\% of the existing contents are cached to fulfill the future needs. While LRU achieves the least cache portion (70\%), Machine Learning approach maintains the medium caching portion as approximately 75\% of the existing media files. The curve of the ML approach also demonstrates that when the number of topics is restricted to 100-150, the ML approach achieves the stable status---increasing of the topic number results in little increase of the contents to be cached. This part results cooperate with the ``hit cache portion" to obtain further conclusions.

\subsubsection{Hit cache portion}
In this section, the hit cache portion is employed as the criterion to evaluate the models. Hit cache portion is portion of how much footprint of media files in the testing dataset is hit by the obtained topics. As we mentioned before, since conventional ``hit rate" measurement is not suitable for Twitter caching scenario, we propose this criterion to demonstrate the coherency between the predicted topics and the future media contents in the testing dataset.

\begin{figure}[htbp]
   \begin{center}
        \includegraphics[width=3.4in]{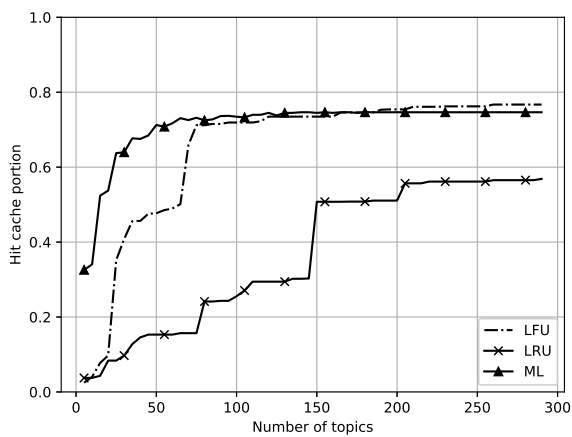}
        \caption{The hit caching portion of the testing dataset (different approaches).}
        \label{Total_size_test}
    \end{center}
\end{figure}

In Fig.\ref{Total_size_test}, the hit cache portion of the three approaches has been demonstrated. Regarding the graph, LFU and ML achieves much better performance (approximately 75\%) compared to the LRU (less than 60\%). This leads to the conclusion that the ML approach and the LFU approach are feasible of associating the obtained topics with the actual caching contents. However, considering the results from the ``cache portion" section, LFU approach is actually costing more footprint to achieve the satisfactory result. This leads to the conclusion that our ML approach caches less redundant contents compared to the conventional LFU method.
The reason for not achieving 100\% caching accuracy is that some caching contents in the testing dataset are irrelevant to the existing media files, while leads to the situation that the topics are not capable of being associated with this part of contents.

To reach a conclusion, the machine learning approach we proposed is evaluated through several different criteria against two conventional caching algorithms. With higher tweet hit rate, our proposed ML approach is capable of achieving satisfying topic prediction results from the tweet text aspect. Regarding to the caching contents evaluation we demonstrate above, the ML approach is feasible of achieving high consistency between the topics and the future caching contents. The cache portion results illustrate that the ML approach achieves less caching redundancy.
\section{Conclusions and Future Work}

In this paper, a novel Twitter aided content caching (TAC) framework, which associated the tweets with the BS information, was proposed. To associate Twitter events with the relative BS, the dataset was established to map tweets to their corresponding BS. Three machine learning approaches of allocating Twitter topics with geographic information of BS were evaluated. Compared to LSTM model with skip-gram embedding, LDA and LDA-based approaches were capable of generating satisfactory predicting results in different regions. Regarding the results and the tendency of perplexity, our novel LSTM model with skip-gram-Geo-aware embedding was quite compatible to process the tweets with BS information. With the aid of machine learning based wireless caching techniques, the redundancy of the caching content was able to be diminished. Combining different types of machine learning approaches, the extracted topics are capable of providing guidance of what text-related contents to cache at the BS. Regarding the different situation of caching at social media platforms and that of traditional media contents providers, such as video sites, specific evaluation criteria are proposed to demonstrate the effectiveness and the accuracy of our proposed framework.

\linespread{0.85}
\bibliographystyle{IEEEtran}
\bibliography{mybib}

\end{document}